\documentclass[useAMS,usenatbib]{mn2e}
\usepackage{graphicx}
\usepackage{rotating}
\usepackage{pdflscape}
\usepackage{natbib}
\usepackage{graphicx}
\usepackage{graphics}
\usepackage{psfrag}
\usepackage{amssymb}
\usepackage{color,soul}
\bibpunct{(}{)}{;}{a}{}{,}

\def\Re{\ensuremath{R_{\oplus}}}
\def\Me{\ensuremath{M_{\oplus}}}
\def\Rt{\ensuremath{R_{\rm T}}}
\def\Rpl{\ensuremath{R_{\rm PL}}}

\def\Mpl{\ensuremath{M_{\rm PL}}}

\def\ergscm{erg\,s$^{-1}$\,cm$^{-2}$}

\def\teq{$T_{\rm eq}$}
\title[]{Identifying the ``true'' radius of the hot sub-Neptune CoRoT-24b by mass loss modelling}
\author[H. Lammer  et al.]{H. Lammer$^{1}$,
			  N. V. Erkaev$^{2,3}$,
			  L. Fossati$^1$,
			  I. Juvan$^1$,
			  P. Odert$^{1}$,
			  P. E. Cubillos$^1$,
	       \newauthor E. Guenther$^4$,
	       		  K. G. Kislyakova$^{1}$,
			  C. P. Johnstone$^5$,
			  T. L\"{u}ftinger$^{5}$,
			  M. G\"{u}del$^5$\\
$^1$Space Research Institute, Austrian Academy of Sciences, Schmiedlstr. 6, A-8042, Graz, Austria\\
$^2$Institute of Computational Modelling SB RAS, 660036, Krasnoyarsk, Russian Federation\\
$^3$Siberian Federal University, Krasnoyarsk, Russian Federation\\
$^4$Th\"{u}ringer Landessternwarte Tautenburg, Sternwarte 5, D-07778 Tautenburg, Germany\\
$^5$Institute for Astronomy, University of Vienna, T\"{u}rkenschanzstrasse 17, 1180 Vienna, Austria}
\begin{document}

\date{}

\pagerange{\pageref{firstpage}--\pageref{lastpage}} \pubyear{2015}

\label{firstpage}

\maketitle

\begin{abstract}
{For the hot exoplanets CoRoT-24b and CoRoT-24c, observations have provided transit radii \Rt\ of $3.7\pm0.4R_{\oplus}$ and $4.9\pm0.5R_{\oplus}$, and masses of $\le5.7M_{\oplus}$ and $28\pm11M_{\oplus}$, respectively. We study their upper atmosphere structure and escape applying an hydrodynamic model. Assuming $R_{\rm T}\approx R_{\rm PL}$, where \Rpl\ is the planetary radius at the pressure of 100\,mbar, we obtained for CoRoT-24b unrealistically high thermally-driven hydrodynamic escape rates. This is due to the planet's high temperature and low gravity, independent of the stellar EUV flux. Such high escape rates could last only for $<$100\,Myr, while \Rpl\ shrinks till the escape rate becomes less than or equal to the maximum possible EUV-driven escape rate. For CoRoT-24b, \Rpl\ must be therefore located at $\approx1.9-2.2R_{\oplus}$ and high altitude hazes/clouds possibly extinct the light at \Rt. Our analysis constraints also the planet's mass to be $5-5.7M_{\oplus}$. For CoRoT-24c, \Rpl\ and \Rt\ lie too close together to be distinguished in the same way. Similar differences between \Rpl\ and \Rt\ may be present also for other hot, low-density sub-Neptunes.}
\end{abstract}
\begin{keywords}
planets and satellites: atmospheres -- stars: ultraviolet -- hydrodynamics
\end{keywords}
\section{Introduction}
The CoRoT satellite detected two Neptune-size planets with transit radii \Rt\ of $\approx$3.7\,\Re\ and $\approx$4.9\,\Re\ with orbital periods of $\approx$5.11 and $\approx$11.76\,days around an old main-sequence K1V star \citep{Alonso2014}. Radial velocity measurements led only to an upper limit of 5.7\,\Me\ for the hotter planet, while for the cooler, more massive planet a mass of 28$\pm$11\,\Me\ was obtained. From the deduced low average densities of $\le0.87$\,g\,cm$^{-3}$ and $1.31\pm0.65$\,g\,cm$^{-3}$ of CoRoT-24b and CoRoT-24c respectively, one can expect that the cores of both planets are surrounded by H$_2$-dominated envelopes. This assumption agrees well with previous studies, concluding that protoplanetary cores with masses $>$1.5\,\Me\ and radii $\ge$1.6\,\Re\ would hardly lose their H$_2$-envelopes under the action of the stellar high-energy flux
\citep{Lammer2014,Erkaev2015,Johnstone2015,Rogers2015,Stoekl2015}.

We aim at understanding whether the low density and large radius of the smaller planet are physically possible. In Sect.~\ref{sec:modelling} and \ref{sec:results} we describe the applied hydrodynamic model, and present the obtained upper atmosphere structure and loss rates. In Sect.~\ref{sec:discussion} we compare the results and discuss the implications of our findings. We conclude in Sect.~\ref{sec:conclusions}.
\section{Modelling approach}\label{sec:modelling}
To study the EUV-heated upper atmosphere structure and hydrogen thermal escape rates, we apply an energy absorption and 1-D hydrodynamic upper
atmosphere model described in detail in \citet{Erkaev2013,Erkaev2015,Erkaev2016} and \citet{Lammer2014}. The model solves the system of hydrodynamic equations for mass,
\begin{equation}
\frac{\partial \left(\rho R^2\right)}{\partial t} + \frac{\partial \left(R^2 \rho v \right)}{\partial R} = 0,
\end{equation}
momentum,
\begin{equation}
\frac{\partial \left(\rho v R^2\right)}{\partial t} + \frac{\partial R^2 \left(\rho v^2 + P\right)}{\partial R}=-\rho R^2 \frac{\partial \phi}{\partial R} + 2 P R,
\end{equation}
and energy conservation,
\begin{eqnarray}
\frac{\partial R^2 \left [\frac{\rho v^2}{2}+ E_{\rm T}\right ]}{\partial t}+
\frac{\partial  R^2 v\left[\frac{\rho v^2}{2} + E_{\rm T} + P \right]}{\partial R}= \nonumber\\
-\rho v \frac{\partial{\phi}}{\partial R}R^2+Q_{\rm EUV}R^2 + \frac{\partial}{\partial R}
\left(\kappa R^2\frac{\partial T}{\partial R}\right).\label{eq:energy_conservation}
\end{eqnarray}
Because the influence of the Roche Lobe can not be neglected, we consider also the energy $\phi$ per unit mass of a test particle in the ecliptic plane
\begin{equation}
\phi=-G\frac{M_{\rm PL}}{R}-G\frac{M_{\rm st}}{d-R}-G\frac{\left(M_{\rm PL} + M_{\rm st}\right)a^2}{2d^3}+const,
\end{equation}
where $G$ is Newton's gravitational constant, \Mpl\ is the planetary mass, $a$ is the distance from a given point to the center of mass, $M_{\rm st}$ is the stellar mass and, $d$ is the distance between the star and the planet. In the equations above, $R$ is the planetocentric distance, $t$ is the time, $T$ the atmospheric temperature, $Q_{\rm EUV}$ is the EUV volume heating rate, $\rho$ the atmospheric mass density, $v$ the gas flow velocity, $P$ the atmospheric pressure, and $\kappa=4.45\times 10^4(T/1000)^{0.7}$ erg\,cm$^{-1}$\,s$^{-1}$\,K$^{-1}$ is the thermal conductivity \citep{Watson1981}. The thermal energy $E_{\rm T}$ of the atmospheric particles per unit volume can then be written as,
\begin{equation}
E_{\rm T}=\left[\frac{3}{2}\left(n_{\rm H} + n_{\rm H^+} + n_{\rm e}\right)+\frac{5}{2}\left(n_{\rm H_2}+n_{\rm H_2^+}\right)\right] k T,
\label{eq:thermalEnergy}
\end{equation}
with $k$ being the Boltzmann constant and $n_{\rm H}$, $n_{\rm H_2}$ neutral particle and $n_{\rm H^+}$, $n_{\rm H_2^+}$ ion densities. The
continuity equations for these number densities are solved as described in \citet{Erkaev2016}. To calculate Q$_{\rm EUV}$, we adopt the average EUV photoabsorption cross sections $\sigma_{\rm EUV}$ of $5\times10^{-18}$\,cm$^{2}$ and $3\times10^{-18}$\,cm$^{2}$ for H atoms and H$_2$ molecules, respectively \citep{Murray2009,Erkaev2013,Lammer2014,Erkaev2015}, in agreement with what is provided by experimental and theoretical data \citep{Bates1962,Cook1964,Beynon1965}.

To estimate the EUV (10-92\,nm) luminosity of CoRoT-24, we consider a stellar rotation period of 27.3\,days \citep{Alonso2014}. Using the scaling laws of \citet{Wright2011} and a $\beta$ \citep[the power-law index correlating the Rossby number at saturation and the X-ray flux; see Eq.\,1 of ][]{Wright2011} of $-$2.18, we obtain $L_{\rm X}=4.51\times10^{27}$\,erg\,s$^{-1}$, hence $L_{\rm EUV}=3.82\times10^{28}$\,erg\,s$^{-1}$ \citep{Sanz2011}. Therefore, we adopted EUV fluxes of 4330\,\ergscm\ and 1410\,\ergscm\ for CoRoT-24b and CoRoT-24c, respectively. For comparison, the X-ray luminosity of HD189733, which has similar parameters to CoRoT-24 is $L_{\rm X}=2.95\times10^{28}$\,erg\,s$^{-1}$ \citep{Schmitt2004}. By using instead the other value of $\beta$ of $-$2.7 given by \citet{Wright2011}, we obtain $L_{\rm X}=1.27\times10^{27}$\,erg\,s$^{-1}$.

The 1-D simulations also include the effects of ionization, dissociation and recombination. The photoionization cross sections, the dissociation and recombination coefficients, the continuity equations for the number densities of the atomic hydrogen atoms and ions, and those for the neutral and ionized molecules were taken from \citet{Yelle2004} and \citet{Erkaev2016} and scaled proportionally to the EUV fluxes derived for each planet. The electron density is determined from the quasi-neutrality condition.

The upper boundary of the simulation domain is set at the Roche lobe of $\approx$24\,\Re\ and $\approx$71.7\,\Re\ for CoRoT-24b and 24c, respectively. The hydrodynamic model can be applied when enough collisions occur, which is the case if Knudsen number $Kn = \lambda/H < 0.1$ \citep{Volkov2011}, where $\lambda$ is the mean free path and $H$ is the local scale height. This Knudsen criterion is fulfilled within the calculation domain. For all calculations we considered a heating efficiency of 15\% \citep{Shematovich2014}.
\begin{figure}
\begin{center}
\includegraphics[width=8.0cm]{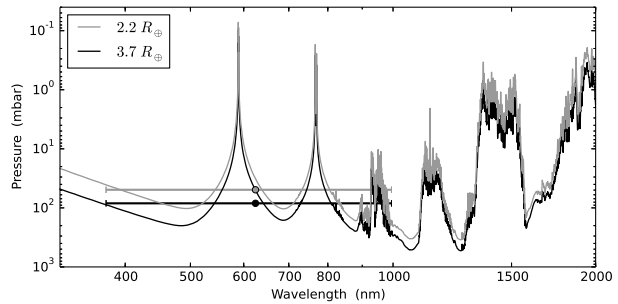}
\caption{Atmospheric pressure as a function of wavelength at which a solar metallicity clear (i.e., without clouds and hazes) atmosphere of a planet becomes opaque, hence where $\tau$\,$\approx$\,1. Calculations have been done for two planetary radii and a planetary mass of 5.7\,\Me. The points show the integral of the convolution of the two model curves with the transmission function of the CoRoT filter at the center of the wavelength span covered by the filter, which is indicated by the horizontal bars.}
\label{fig:tau}
\end{center}
\end{figure}
We set the lower boundary of the simulations at $R_{\rm PL}$, where we assume a temperature equal to the planetary equilibrium temperature \teq. Here we also assume an H$_2$ number density corresponding to an atmospheric pressure of 100\,mbar, which is near the level where the tangential optical depth $\tau$\,=\,1 for most transparent continuum bands \citep{Brown2001,Lopez2014}.

To prove the validity of this assumption we use a radiative-transfer transmission-spectrum model\footnote{{\tt https://github.com/exosports/BART}} for a solar composition atmosphere and the $T$-$P$ profiles obtained from the hydrodynamic code extended with isothermal hydrostatic profiles at high pressures, assuming two different planetary radii (see Sect.~\ref{sec:results}). The main contribution to the absorption comes from H$_2$ Rayleigh scattering ($<$600\,nm), alkali lines (Na and K; 500--800\,nm), H$_2$-H$_2$ collision-induced absorption ($>$600\,nm), and H$_2$O ($\approx$900\,nm). CoRoT's observing band covers the range 350--1000\,nm, with the effective wavelength at $\approx$690\,nm. Figure~\ref{fig:tau} shows that $\tau$\,$\approx$\,1 lies at a pressure of $\approx$60--70\,mbar, but this is for a slant geometry  (transmission) and therefore in a radial geometry the pressure at \Rt\ is slightly higher than the average pressure probed by transmission spectroscopy. Differences of a factor of 10 in metallicity do not affect this result.

We run a consistency check of the adopted hydrodynamic code by comparing the calculated H loss rates with those of the well studied hot-Jupiter HD\,209458b, for which upper atmosphere hydrodynamic modelling has been performed by several groups. If we assume similar planetary and atmospheric parameters at the lower boundary \citep{Koskinen2013} we obtain a mass-loss rate of $\approx$7$\times$10$^{10}$\,g\,s$^{-1}$, which agrees well with what previously derived by several other groups \citep[4--7$\times$10$^{10}$\,g\,s$^{-1}$;][]{Yelle2004,Munoz2007,Koskinen2013,Khodachenko2015}.
\section{Results}\label{sec:results}
\subsection{Thermal outflow}
For the first set of calculations, we assumed that \Rt\,=\,\Rpl\,=\,3.7\,\Re\ for CoRoT-24b and 4.9\,\Re\ for CoRoT-24c. The equilibrium temperatures \teq\ of CoRoT-24b and 24c are 1070$\pm$140\,K and 850$\pm$80\,K, respectively \citep{Alonso2014}. Because these temperatures are cooler than the H$_2$ thermal dissociation temperature ($\approx$2000\,K), H$_2$ is the dominant species near the 100\,mbar level. Above the altitude level where dissociation takes place, H atoms become the main species and can escape from the planet's gravitational well. The solutions of our model runs are quasi-stationary with slow time variations of the atmospheric parameters.

For CoRoT-24b and CoRoT-24c, the hydrodynamic model gives H escape rates of $\approx$2.0$\times$10$^{36}$\,s$^{-1}$ (3.35$\times$10$^{12}$\,g\,s$^{-1}$) and $\approx$8.0$\times$10$^{33}$\,s$^{-1}$ (1.34$\times$10$^{10}$\,g\,s$^{-1}$), respectively. Under the assumption that the 100\,mbar level lies near the observed transit radius, the high thermally driven hydrodynamic outflow obtained for CoRoT-24b is caused by the combination of the planet's high temperature and low gravity, independent of the stellar EUV flux. This happens because for hot, low-gravity planets the thermal energy $E_{\rm T}$ shown in Eq.~\ref{eq:thermalEnergy} for a large fraction of the particles in the upper atmosphere is of the order of or larger than the potential energy. As a consequence, the atmosphere flows upwards, even without the need of an additional external energy source, such as the stellar EUV flux. In this case, the stellar EUV flux is not the main driver of the atmospheric escape. This is similar to the so called ``boil-off'' evaporation regime of very young planets \citep{Owen2016}.
\begin{table}
\renewcommand{\baselinestretch}{1}
\caption{Results of the hydrodynamic simulations for CoRoT-24b with \Mpl\,=\,5.7\,\Me, \teq\,=\,1070\,K, and \Rpl\,=\,R$_{\rm 100\,mbar}$ ranging between 1.7 and 3.7\,\Re. The table lists the planetary densities $\rho_{\rm PL}$ in g\,cm$^{-3}$, hydrodynamic $L_{\rm HY}$ and energy-limited $L_{\rm EN}$ escape rates in s$^{-1}$ and their ratio ($L_{\rm HY}/L_{\rm EN}$).}
\label{tab:table}
\begin{center}
\begin{tabular}{ccccc}
\hline
\hline
$R_{\rm PL}$ & $\rho_{\rm PL}$ & $L_{\rm HY}$ [s$^{-1}$] & $L_{\rm EN}$
[s$^{-1}$] & $L_{\rm HY}/L_{\rm EN}$\\
\hline
3.7 & 0.62 & $2.0\times10^{36}$ & $5.3\times10^{34}$ & 37.7 \\
3.5 & 0.73 & $1.3\times10^{36}$ & $4.8\times10^{34}$ & 27.1 \\
3.0 & 1.16 & $1.8\times10^{35}$ & $1.9\times10^{34}$ &  9.5 \\
2.5 & 2.01 & $2.6\times10^{34}$ & $1.0\times10^{34}$ &  2.6 \\
2.2 & 2.94 & $6.0\times10^{33}$ & $6.0\times10^{33}$ &  1.0 \\
2.0 & 3.92 & $2.3\times10^{33}$ & $5.0\times10^{33}$ & 0.46 \\
1.7 & 6.38 & $4.4\times10^{32}$ & $2.7\times10^{33}$ & 0.16 \\
\hline
\hline
\end{tabular}
\end{center}
\end{table}

The high escape rate obtained for CoRoT-24b could in principle last for $<$100\,Myr (see Sect.~\ref{sec:indentifying}), which is inconsistent with the old age of the system: with the exception of very young planets, the escape rate can only be less than or equal to the maximum possible EUV-driven H escape rate, which can be analytically estimated with the energy-limited formula \citep[e.g.,][]{Watson1981,Erkaev2007,Lammer2009,Luger2015},
\begin{equation}
L_{\rm EN}=\frac{\pi\eta R_{\rm PL}R_{\rm EUV}^2I_{\rm EUV}}{GM_{\rm PL}m_{H}K},
\label{eq:energyLimited}
\end{equation}
\noindent where $\eta$ is the heating efficiency, $I_{\rm EUV}$ is the stellar EUV flux, $R_{\rm EUV}$ is the effective radius where the EUV energy is absorbed in the upper atmosphere \citep{Erkaev2013,Erkaev2015,Erkaev2016}, and $m_{\rm H}$ the mass of the H atom. The factor $K$ takes into account Roche lobe effects \citep{Erkaev2007}.

By applying this formula to CoRoT-24b and CoRoT-24c, we obtain escape rates of 5.3$\times$10$^{34}$\,s$^{-1}$ (8.9$\times$10$^{10}$\,g\,s$^{-1}$) and 5.5$\times$10$^{33}$\,s$^{-1}$ (9.2$\times$10$^{9}$\,g\,s$^{-1}$), respectively. For the hotter and less massive sub-Neptune CoRoT-24b, the escape rate based on the hydrodynamic model $L_{\rm HY}$ is $\approx$40 times higher than $L_{\rm EN}$ obtained from Eq.~\ref{eq:energyLimited}, while for the more massive and cooler planet, CoRoT-24c, $L_{\rm HY}$ is consistent with $L_{\rm EN}$, within the expected uncertainties.

From these estimates it follows that the high escape rate obtained from the hydrodynamic simulations for CoRoT-24b cannot be realistic. The only possible solution to this discrepancy is that \Rt\ is not close to the 100\,mbar level, but lies much higher in the atmosphere.
\begin{figure}
\begin{center}
\vspace{-0.6cm}
\includegraphics[width=8.5cm,clip]{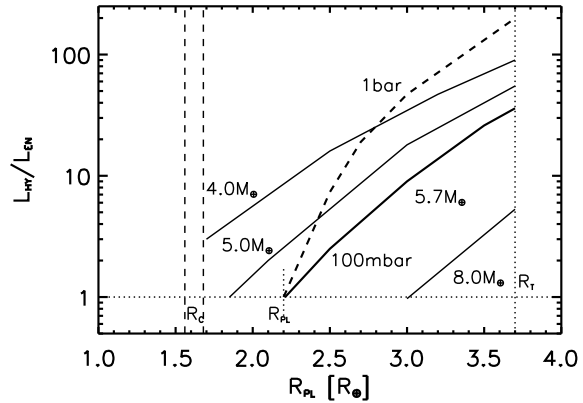}
\vspace{-0.6cm}
\caption{Ratio between hydrodynamic escape rates $L_{\rm HY}$ and corresponding energy-limited escape rates $L_{\rm EN}$ as a function of planet radius, considered to be at the 100\,mbar level, and planet mass. The dotted horizontal line indicates $L_{\rm HY}$\,=\,$L_{\rm EN}$. The vertical lines mark the transit radius, and the possible range of planet (at the 100\,mbar pressure level) and core radii $R_{\rm c}$ (see Sect.~\ref{sec:discussion}) assuming a mass of 5.7\,\Me. The dashed line indicates the results obtained assuming a mass of 5.7\,\Me\ and a pressure at \Rpl\ of 1\,bar.}
\label{fig:LhyvsLen}
\end{center}
\end{figure}
\subsection{Identifying the 100\,mbar level}\label{sec:indentifying}
To find out which set of planet parameters satisfies the $L_{\rm HY}/L_{\rm EN}$\,$\leq$\,1 condition for CoRoT-24b, we run a grid of hydrodynamic models with mass and radius values in the 4.0--8.0\,\Me\ and 1.7--3.7\,\Re\ range, respectively, to directly compare $L_{\rm HY}$ and $L_{\rm EN}$ as a function of planet parameters. Table~\ref{tab:table} shows the results of the hydrodynamic model for CoRoT-24b with \Mpl\,=\,5.7\,\Me\ and \Rpl\ ranging between 1.7 and 3.7\,\Re.

Figure~\ref{fig:LhyvsLen} shows the $L_{\rm HY}/L_{\rm EN}$ ratio as a function of planetary radius and mass and indicates that, assuming \Mpl\,$\leq$\,5.7\,\Me, the condition $L_{\rm HY}/L_{\rm EN}$\,$\leq$\,1 is reached for planetary radii smaller than $\approx$2.2\,\Re. In this work, the transit radius of CoRoT-24b is the main observational uncertainty, but it has however no impact on the results: the $L_{\rm HY}/L_{\rm EN}$\,$\leq$\,1 relation is fulfilled for \Rpl\,$<$\,2.2\,\Re, regardless of the observed radius and its uncertainty. Our simulations at larger masses show that the transit radius lies close to the 100\,mbar level for \Mpl\,$>$\,8\,\Me.

According to formation and structure models of hot (\teq\,$\approx$\,1000\,K) sub-Neptunes \citep[e.g.,][]{Rogers2011}, the transit radius would correspond to an atmospheric mass fraction $f$ of $\approx$0.04, hence an atmospheric mass $M_{\rm AT}$ of $\approx$1.36$\times$10$^{27}$\,g. By assuming a core mass of 5.7\,\Me\ for CoRoT-24b, from the escape rates given in Table~1, one can also roughly estimate the evolution of atmospheric mass with time. Atmospheric masses corresponding to the radii given in Table~1 were estimated from the 1000\,K structure models of \citet{Rogers2011}. As one can see in Fig.~\ref{fig:evolution}, an atmospheric mass above 2.2\,\Re\ corresponding to a radius of 3.7\,\Re\ would be lost within 100\,Myr due to the high thermally-driven escape rates, in agreement with the results of \citet{Owen2016}. Because of the short time period in which the mass-loss rates are exceptionally high and the corresponding fast shrinking of the planet radius, the 100\,mbar level of CoRoT-24b must be located much deeper in the atmosphere, compared to the position of \Rt.
\section{Discussion}\label{sec:discussion}
\begin{figure}
\begin{center}
\vspace{-0.6cm}
\includegraphics[width=8.5cm,clip]{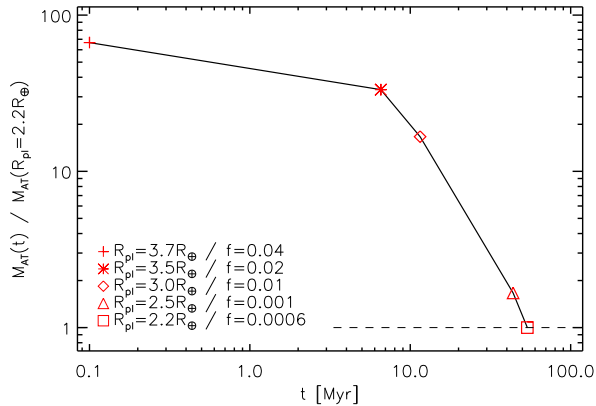}
\vspace{-0.6cm}
\caption{Atmospheric mass $M_{\rm AT}$ evolution normalized to the atmospheric mass corresponding to $R_{\rm PL}=2.2R_{\rm oplus}$ estimated from the results shown in Table 1. The horizontal line indicates $M_{\rm AT}$\,=\,$M_{\rm AT}(2.2R_{\rm PL})$.}
\label{fig:evolution}
\end{center}
\end{figure}
Thermally driven hydrodynamical mass loss could occur at the first stages of planetary evolution \citep{Owen2016}, but is unlikely to occur for older planets because, as shown in Fig.~\ref{fig:evolution}, the atmosphere structure would be modified in a short time. It is also not possible that CoRoT-24b is the remnant of a more massive planet (e.g., hot-Jupiter or hot-Saturn), because the escape rates of such massive objects are not high enough to affect significantly the planet's mass over the main-sequence lifetime of the host star \citep{Yelle2004}.

Both escape rates, i.e. $L_{\rm HY}$ and $L_{\rm EN}$, depend on the adopted EUV flux and heating efficiency $\eta$. For this reason we tested the effect of a higher/lower EUV flux or $\eta$ on the results. A different EUV flux affects both $L_{\rm HY}$ and $L_{\rm EN}$ rates in a similar, though not identical, way. A variation of a factor of $\approx$2 in the EUV flux would lead to a variation on the $L_{\rm HY}/L_{\rm EN}$ ratio of $\approx$0.3, which in terms of planetary radius corresponds to $<$\,0.1\,\Re. Variations in $\eta$ also affect both escape rates. In this case we estimate that a factor of $\approx$2 difference in the heating efficiency would lead to a variation on the planetary radius of $<$\,0.1\,\Re. From the comparison of the escape rates of the hot-Jupiter HD\,209458b described in Sect.~\ref{sec:modelling} we expect that the use of a different hydrodynamic code would lead to mass-loss rates consistent with those presented here within a factor of $\approx$2, which, in terms of the derived planetary radius of CoRoT-24b, corresponds to $\approx$\,0.1\,\Re. We further tested the effect of varying \teq\ \citep[1070$\pm$140\,K;][]{Alonso2014}, finding that variations in \teq\ up to 200\,K have a negligible impact on the results. 
\begin{figure}
\begin{center}
\vspace{-0.6cm}
\includegraphics[width=8.5cm,clip]{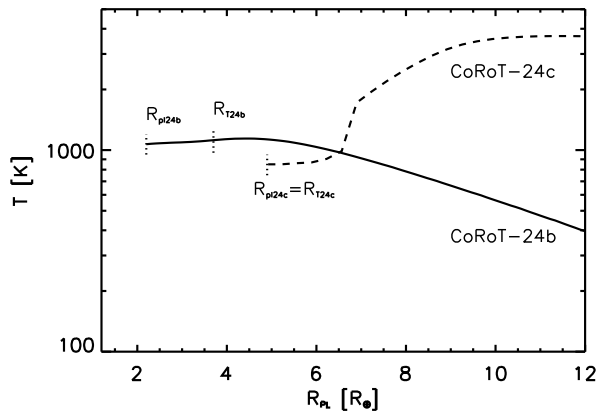}
\includegraphics[width=8.5cm,clip]{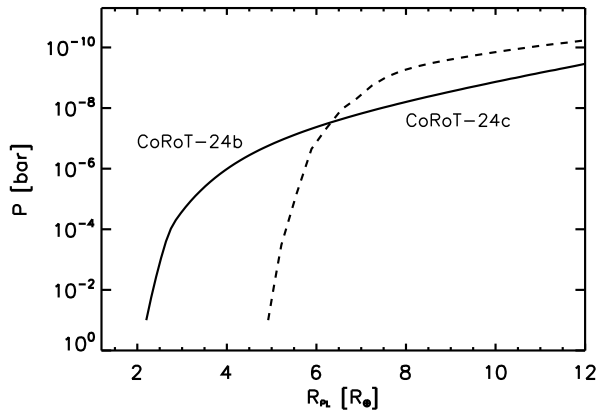}
\vspace{-0.6cm}
\caption{Top: temperature profiles of CoRoT-24b, calculated assuming \Rpl\,=\,2.2\,\Re\ and \Mpl\,=\,5.7\,\Me\ (solid line), and CoRoT-24c (dashed line) as a function of distance in units of \Re. Bottom: same as the top panel, but for the atmospheric pressure.}
\label{fig:profiles}
\end{center}
\end{figure}

On the basis of planetary formation models calculated for planets in the 1--10\,\Me\ range, one can estimate the core radius $R_{\rm c}$ of a planet on the basis of its mass if $M_{\rm PL}\approx M_{\rm c}$ as $R_{\rm c} \approx M_{\rm PL}^{0.27-0.3}$, depending upon the exact composition \citep{Valencia2006}. Taking into account that \Rpl, hence the 100\,mbar level, cannot be too close to $R_{\rm c}$ (the planet needs sufficient atmosphere to produce the observed \Rt), from Fig.~\ref{fig:LhyvsLen} it follows that CoRoT-24b should have a minimum mass of $\approx$5.0\,\Me, with a core radius in the $\approx$1.55--1.68\,\Re\ range and \Rpl\ at the 100\,mbar level in the $\approx$1.9--2.2\,\Re\ range. These values yield an average density of 2.9--4.6\,g\,cm$^{-3}$. The mass contained between \Rpl\ and \Rt\ is $7.5\times10^{-5}$ of \Mpl. We conclude therefore that, accounting for the uncertainties on \Rt, the ``true'' radius $R_{\rm pl}$ of CoRoT-24b is $\approx$30-60\% smaller than the one measured by the transit. For the solar system gas giant planets, \Rpl\ is by definition set at the 1\,bar pressure level. By setting this pressure level at \Rpl\ and following the same procedure described in Sect.~\ref{sec:indentifying}, we obtained again that \Rpl\ would be about 2.2\,\Re\  (see Fig.~\ref{fig:LhyvsLen}).

Figure~\ref{fig:profiles} compares the temperature and pressure profiles of CoRoT-24b, considering a \Rpl\ of 2.2\,\Re\ and a \Mpl\ of 5.7\,\Me, and CoRoT-24c. In the case of CoRoT-24b, the atmosphere is hydrostatic near \Rpl, while it becomes hydrodynamic and cools adiabatically above 5\,\Re. For the cooler and more massive planet, CoRoT-24c, the atmosphere is hydrostatic above \Rt\ and the temperature increases up to 3000\,K due to the EUV heating. Above 10\,\Re\ the atmosphere cools adiabatically. For CoRoT-24b, Fig.~\ref{fig:profiles} shows that assuming \Rpl\,=\,2.2\,\Re\ and \Mpl\,=\,5.7\,\Me, \Rt\ lies near the 1--10\,$\mu$bar pressure level. As one can see from Fig.~\ref{fig:tau}, the absorption at the level of the transit radius cannot be caused by H$_2$ Rayleigh scattering and the alkali lines are too narrow to produce broad-band absorption. At such a high altitude, absorption could be for example produced by clouds, though cloud formation models based on TiO$_2$ calculated for HD\,189733b (\teq\ similar to that of CoRoT-24b) suggest formation pressures of about 0.1\,mbar \citep{Lee2015}.
\section{Conclusions}\label{sec:conclusions}
By comparing the escape rates obtained from a hydrodynamic code with the maximum possible EUV-driven escape rate calculated from the energy-limited formula, we discovered that for the H$_{2}$-dominated close-in sub-Neptune CoRoT-24b the planetary radius, by definition set at the 100\,mbar pressure level, is 30-60\% smaller than the observed transit radius. Additional considerations on the range of possible planetary core radii allowed us to constrain the most likely values of the planetary mass and radius to lie in the range 5.0--5.7\,\Me\ and 1.9--2.2\,\Re, respectively.

The Kepler satellite has discovered several similar, low-density ``sub-Neptunes'' and it is very likely that the transit radii measured for the vast majority of these planets also differs from the true planetary radius, therefore introducing a systematic bias in the measured radii. This has dramatic implications for example in the determination of the mass-radius relation \citep[e.g.,][]{Petigura2013,Weiss2014,Lopez2014} and for planet population synthesis studies \citep[e.g.,][]{Beauge2013}. Our results may also help explaining some of the discrepancies discovered between the planet radius distribution observed by Kepler and that derived from population synthesis models \citep{Jin2014}. This has also profound implications for our understanding of the structure of the large number of transiting low-mass exoplanets expected to be observed and discovered in the near future by the CHEOPS, TESS, and PLATO space missions and characterized by JWST.

\section*{Acknowledgments}
HL and IJ acknowledge the funding by FFG-847963. TL acknowledges funding by FFG-ASAP11. The authors thank the FWF NFN project S11601-N16 and the subprojects S11604-N16 and S11607-N16. HL, NVE, and PO acknowledge the FWF project P27256-N27. NVE acknowledges support by the RFBR grant No.\,15-05-00879-a. The authors thank the anonymous referee for the useful comments that led to a considerable improvement of the manuscript. We thank R. Alonso for fruitful discussions.

\label{lastpage}

\end{document}